\documentclass[%
 reprint,
superscriptaddress,
 amsmath,amssymb,
 aps,
 prd,
]{revtex4-2}

\usepackage{graphicx}
\graphicspath{{./figures/}}
\usepackage{xcolor}
\usepackage{ulem}
\usepackage{dcolumn}
\usepackage{bm}
\usepackage{hyperref}
\hypersetup{colorlinks, linkcolor = [rgb]{0, 0, 0.5}, citecolor = [rgb]{0,0.0,0.5}, urlcolor = [rgb]{0,0.0,0.5}}
\usepackage[caption=false]{subfig}
\usepackage{siunitx}
\usepackage[capitalise]{cleveref}
\allowdisplaybreaks

\usepackage{graphicx}
\usepackage{dcolumn}
\usepackage{bm}
\usepackage{textgreek}
\usepackage{xcolor}

\begin{document}

\preprint{APS/123-QED}

\title{Study of charm hadronization and in-medium modification at the Electron-ion Collider in China}

\author{Senjie Zhu}
\email{zhusenjie@mail.ustc.edu.cn}
\affiliation{University of Science and Technology of China, Hefei, Anhui Province 230026, China}

\author{Xiao Huang}
\affiliation{University of Science and Technology of China, Hefei, Anhui Province 230026, China}

\author{Lei Xia}
\affiliation{University of Science and Technology of China, Hefei, Anhui Province 230026, China}

\author{Aiqiang Guo}
\affiliation{Institute of Modern Physics, Chinese Academy of Sciences, Lanzhou, Gansu Province 730000, China}
\affiliation{University of Chinese Academy of Sciences, Beijing 100049, China}

\author{Yutie Liang}
\affiliation{Institute of Modern Physics, Chinese Academy of Sciences, Lanzhou, Gansu Province 730000, China}
\affiliation{University of Chinese Academy of Sciences, Beijing 100049, China}
\affiliation{Guangdong Provincial Key Laboratory of Nuclear Science,\\
Institute of Quantum Matter, South China Normal University, Guangzhou 510006, China}

\author{Yifei Zhang}
\email{ephy@ustc.edu.cn}
\affiliation{University of Science and Technology of China, Hefei, Anhui Province 230026, China}

\author{Yuxiang Zhao}
\affiliation{Institute of Modern Physics, Chinese Academy of Sciences, Lanzhou, Gansu Province 730000, China}
\affiliation{Southern Center for Nuclear-Science Theory (SCNT), Institute of Modern Physics,
Chinese Academy of Sciences, Huizhou, Guangdong Province 516000, China}
\affiliation{University of Chinese Academy of Sciences, Beijing 100049, China}
\affiliation{Key Laboratory of Quark and Lepton Physics (MOE) and Institute of Particle Physics,\\
Central China Normal University, Wuhan 430079, China}


\date{\today}

\begin{abstract}
Charm quark production and its hadronization in $ep$ and $eA$ collisions at the future Electron-Ion Collider in China (EicC) will help us understand the quark/gluon fragmentation processes and the hadronization mechanisms in the nuclear medium, especially within a poorly constrained kinematic region ($x<0.1$). In this paper, we report a study on the production of charmed hadrons, $D^0$ and $\Lambda_c^+$, reconstructed with a dedicated \textsc{geant4} simulation of vertex$\,\&\,$tracking detectors designed for EicC. The $\Lambda_c^+$/$D^0$ ratios as functions of multiplicity and $p_T$, as well as the $D^0$ double ratio are presented with projected statistical precision.
\end{abstract}

\maketitle

\section{Introduction}
Exploring the most fundamental building blocks of the universe is the paramount objective of modern physics. 
Over the past five decades, we've come to understand that matter, in its essence, is composed of nuclei and nucleons. These, in turn, are made up of even more basic components known as quarks, which are confined in a colorless bound state through the exchange of gluons. This interaction among quarks and gluons is governed by the theory of strong force, known as Quantum Chromodynamics (QCD).
However, up to now, the constituent interactions and the dynamical distributions of quarks and gluons inside nucleons, in particular in the kinematic regime dominated by sea quarks and gluons, are still unclear or poorly constrained by existing experiments~\cite{nnpdf,nr_pdf}. The next-generation experimental facilities with electron-ion collision have been proposed for the quantitative study of matter in this new regime, such as Electron-Ion Collider at BNL (EIC-BNL)~\cite{eic_white_Paper,yellow_report_eic} and Electron-ion collider in China (EicC)~\cite{eicc_white_paper}.


EicC is proposed to operate at a center-of-mass (c.m.) energy ranging from 15 GeV to 20 GeV, with a luminosity of approximately 
$2.0 ~\times ~10^{33}$ cm$^{-2}\cdot$s$^{-1}$~\cite{eicc_white_paper}. This would enable EicC to explore the kinematic region dominated by sea quarks, effectively bridging the gap between the EIC-BNL and JLab 12 GeV experiments. The primary objective of EicC is to delve into the partonic structure and three-dimensional tomography of nucleons and nuclei, the
cold nuclear matter effects, as well as the origin of proton mass.
This ambitious endeavor aims to significantly advance our understanding of these fundamental aspects of nuclear physics.


Understanding the interactions of partons with nuclear matter, as well as their fragmentation~\cite{lund_model} or combination~\cite{coalescence_hot_medium} to form hadrons—a process known as hadronization—is a fundamental question that future EicC experiments aim to answer. Heavy quarks, due to their large masses, are predominantly produced in the initial hard scatterings of collisions and undergo the complete evolution of the nuclear medium system. This makes them an ideal probe for measuring nuclear medium effects and heavy quark hadronizations in high-energy heavy-ion collisions~\cite{prompt_d_pbpb_alice,open_charm_production_star,prompt_d0_ppb_alice}.

For example, the enhancement of $\Lambda_{c}^+/D^{0}$ ratios in heavy-ion collisions compared with elementary particle collisions has been measured by STAR~\cite{open_charm_production_star} and ALICE~\cite{lhc_frag_18,lc2d0_ratio_pbpb_alice}, revealing the hadronization mechanism of charm quark with light quarks in hot ~\cite{coalescence_hot_medium,tamv} and dense medium~\cite{helac}. 
However, in these collisions, it is a complex task to separate the effect of the cold nuclear medium (CNM) from the dominant medium response caused by the hot quark-gluon plasma (QGP). This is where electron-ion collisions come into play, offering an ideal platform to study the CNM effects~\cite{sidis_nuclei_hermes} in a more transparent system where it is believed no QGP is formed. Moreover, two theories with different time scales - parton energy loss and the hadron absorption model~\cite{hadron_absorption} - can successfully describe the suppression of light hadrons in electron-ion collisions. The measurement of heavy quark production presents a novel opportunity to reveal the mechanisms of hadronization. To address these inquiries, it is essential to investigate charm hadronization and the production of open charmed hadrons in future EicC experiments.

In this paper, we perform a comprehensive simulation study with a \textsc{geant4}~\cite{geant4} detector configuration for open charm hadron production. The report is organized as follows. Section II introduces the simulation setup and process. Section III presents the simulation results and a summary is given in Section IV.

\section{Simulation Setup}
\hspace{\parindent}
This study employs the \textsc{pythia} event generator, as referenced in~\cite{pythia6}. Two distinct versions of this generator, namely pythiaeRHIC (\textsc{pythia 6.4}) and \textsc{pythia 8.3}, are utilized in our analysis. The configuration for \textsc{pythia 6.4} is detailed extensively in Ref.~\cite{PYTHIA6setting}. The physical processes including vector-meson diffractive and resolved processes, semihard QCD $2\rightarrow 2$ scattering, neutral boson scattering off heavy quarks within the proton, and photon-gluon fusion, are turned on. Several alternative hadronization models in \textsc{pythia} 8.3 are used for charm hadronization studies, e.g., implemented color reconnection models. The kinematic variables used are listed in Table \ref{tab:var_def}.\par

\begin{table*}[t]
\caption{\label{tab:var_def}%
Kinematic variable definition.
}
\begin{ruledtabular}
\begin{tabular}{p{3cm} p{16cm}}

$k$ and $k'$& The four-momentum of the incoming electron and the scattered electron\\
$p$ and $p_h$& The four-momentum of the incoming proton and the produced hadron\\
$q=k-k'$& The four-momentum of emitted virtual photon \\
$Q^2=-q^2$ & The negative square invariant momentum transfer of emitted virtual photon\\
$x_B=Q^2/(2p\cdot q)$&  Bjorken scaling variable\\
$y=p\cdot q/(p\cdot k)$&  The fraction of the incoming electron's energy transferred to the hadronic system\\
$z=p\cdot p_h/(p\cdot q)$& The momentum fraction of the virtual photon to be carried by the produced hadron\\
$W=\sqrt{(q+p)^2}$ & The center-of-mass energy of $\gamma^*$-Nucleon system \\
$\nu=q\cdot p/m_N$ & The energy of $\gamma^*$ in nucleon rest frame ($m_N$ is the mass of the nucleon)\\
 
\end{tabular}
\end{ruledtabular}
\end{table*}

\begin{figure}[htp]
  \centering
  \includegraphics[width=0.45\textwidth]{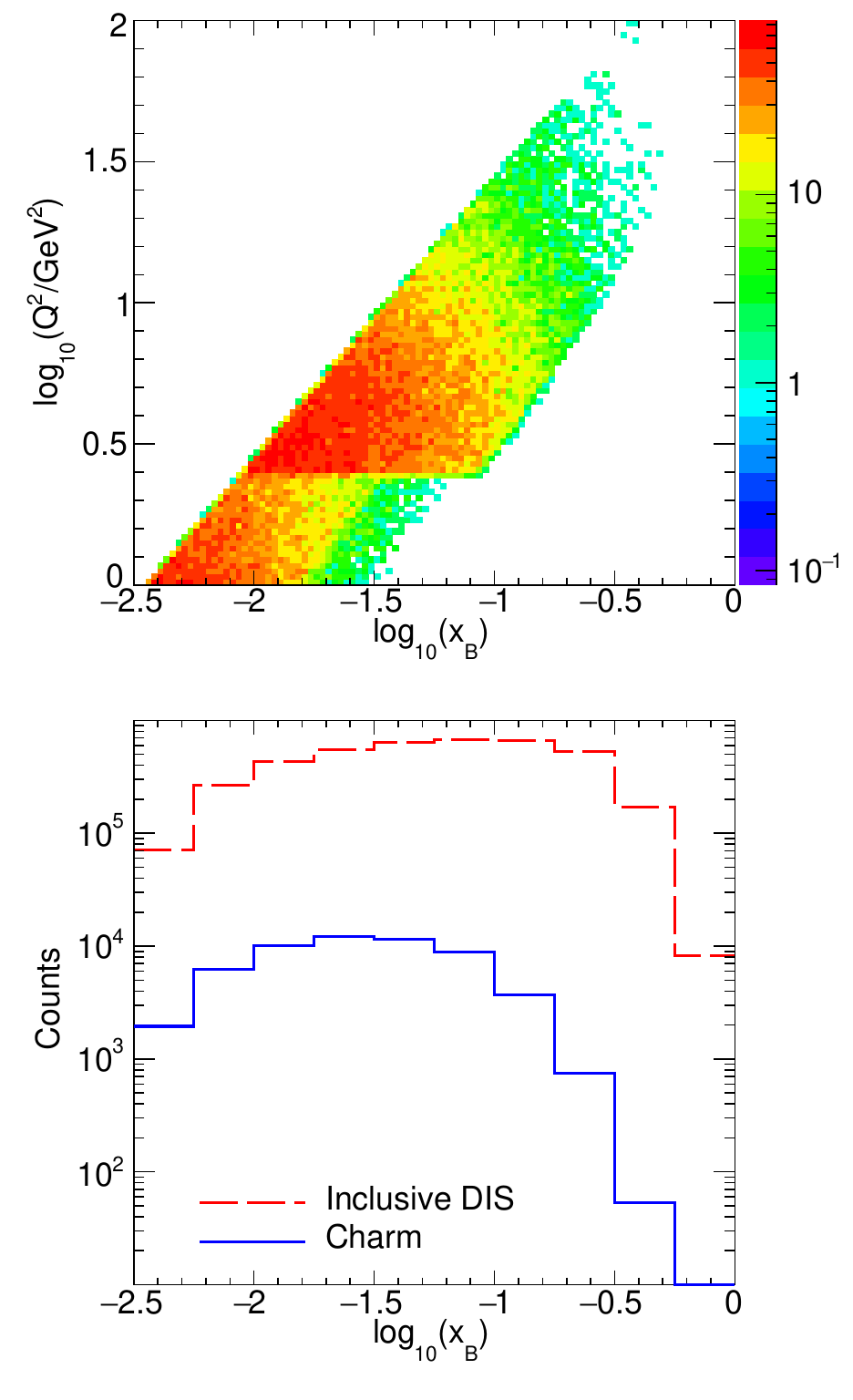}
  \caption{\label{fig:minbias_kinematic_coverage} (Top) $\rm{log}_{10}$$(x_B)-\rm{log}_{10}$$(Q^2/GeV^2)$ distribution of charmed events per $1\,\rm{fb}^{-1}$. (Bottom) $\rm {log}_{10}$$(x_B)$ distribution of inclusive DIS and charm per $1\,\rm{fb}^{-1}$ ($\rm Q^2> 1\,GeV^2$ ).}
\end{figure}

\begin{figure}[htp]
  \centering
  \includegraphics[width=0.4\textwidth]{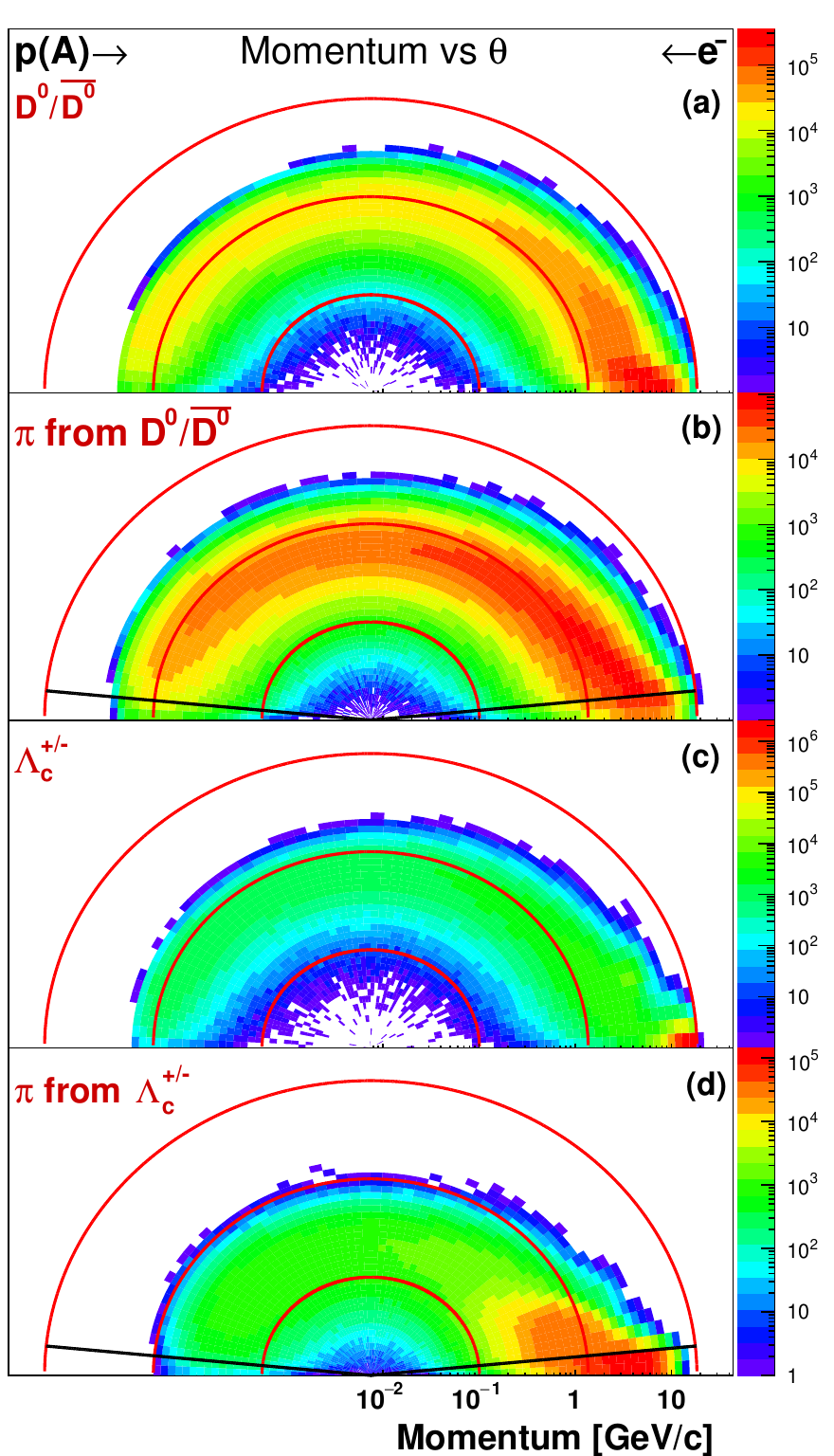}
  \caption{\label{fig:d0_decay_kinematic} The momentum-$\rm \theta$ distributions of the particles produced by \textsc{pythia}. The particle types are labeled respectively in the panel for $D^0/\bar{D^0}$, $\Lambda_c^{\pm}$ and pions from their decay.}
\end{figure}

The kinematic coverage of  the collision with electron (3.5\,GeV) and proton (20\,GeV) is shown in Fig.\ref{fig:minbias_kinematic_coverage} (top panel), which presents the $\rm{log}_{10}$$(x_{B})-\rm{log_{10}}(Q^2)$ distribution of charmed events per $1\,\rm{fb}^{-1}$ integral luminosity. As shown in the $x_{B}-Q^2$ distribution, one step along $Q^2$ distribution exists at $\rm{log}_{10}($$Q^2$$)=0.4$. 
The reason is that the subprocess cross-section ($\gamma^*q\rightarrow q$) in \textsc{pythia 6} is deliberately set to zero when the photon's virtuality $Q^2$ approaches zero. This is done to prevent double-counting with real-photon physics processes\cite{pythia6}.
In the above beam setting and with DIS requirement $Q^2>1\,$GeV$^2$, the $x_{B}$ distributions are shown in the bottom panel of Fig.\ref{fig:minbias_kinematic_coverage}, including inclusive DIS events and charmed events. Charmed events account for $1.4\%$ of inclusive DIS events. The momentum-$\theta$ phase space 2-dimension distribution of $D^0$ is shown in Fig.\ref{fig:d0_decay_kinematic} (the first panel). 
Due to the higher momentum of the beam protons compared to the beam electron, the $D^0/\bar{D^0}$ distribution is skewed towards the direction of the proton beam. Consequently, as illustrated in the second panel of Fig.\ref{fig:d0_decay_kinematic}, the pions from $D^0\rightarrow \pi K$ decay are given a boost, causing them to be produced more in the forward direction of the proton beam.
As illustrated in Fig.\ref{fig:d0_decay_kinematic} for $\Lambda_c^{\pm}$, the distributions of $\Lambda_c^{\pm}$ and the pions from its decay are even more skewed towards the direction of the proton beam. There is a region near $\theta=0$ where the number of $\Lambda_{c}^{\pm}$ is significantly larger than in other areas, and the $\Lambda_c^{\pm}$ in this region has much higher energy. 
Furthermore, $\Lambda_c^+$ is much more abundant than $\Lambda_c^-$ in this region. This phenomenon is caused by the formation of beam remnants~\cite{beam_remnants}, a charm quark produced via the hard scattering and $u$, $d$ quarks from the beam proton form a $\Lambda_c^{+}$ together. Because the $u$, $d$ quarks from protons carry a large fraction of the proton momentum, the beam remnant $\Lambda_c^+$ is very energetic. \par

The specific design and layout for EicC detectors have been detailed in Ref.~\cite{anderle2023probing}, which include a tracking and vertex detector system. This report primarily focuses on the tracking and vertex detectors.
The detector acceptance is $-3<\eta<3$ ($5.7^{\circ} \sim 174.3^{\circ}$). In Fig.\ref{fig:d0_decay_kinematic} (panel (b) and (c)), the acceptance region is marked by two black lines. In Ref.~\cite{anderle2023probing}, the performances have been obtained through \textsc{geant4} simulation for different particle species ($e,\mu,\pi,K,p$) under a $1.5\,$T magnetic field configuration. 

\begin{figure*}[htp] 
  \centering
  \includegraphics[width=0.95\textwidth]{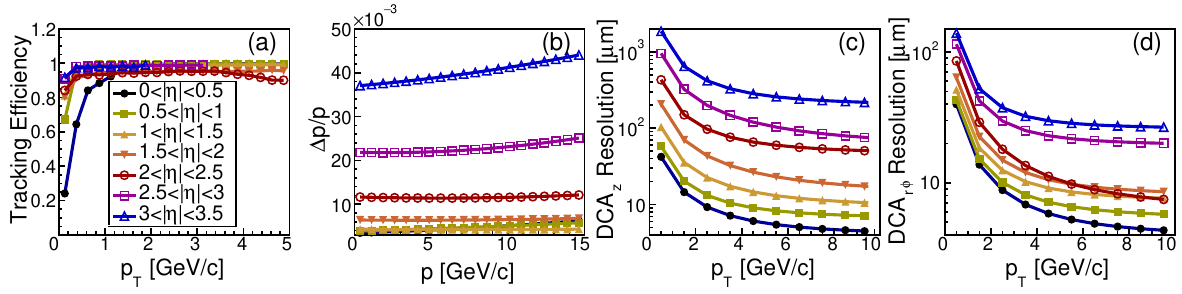}
  \caption{\label{fig:track_vertex_detector_performance} The detector's performance for $\pi$: (a) the efficiency of track reconstruction, (b) the resolution of the reconstructed track momentum, (c) the resolution of the reconstructed track $DCA$ in $z$ direction, (d) the resolution of the reconstructed track $DCA$ in the $r\phi$ plane. The different marker styles and line colors of the lines represent the different $\eta$ regions of $\pi$, which are shown in the legend of the panel (a).}
\end{figure*}


The performances for $\pi$ are shown as an example in Fig.\ref{fig:track_vertex_detector_performance}. In this study, the Distance of the Closest Approach ($DCA$) are defined as the closest distance between the reconstructed track and the primary vertex in $r\phi$ plane ($DCA_{r\phi}$) and $z$-axis ($DCA_z$). The tracking efficiency, the momentum resolution, the $DCA_z$ and $DCA_{r\phi}$ resolutions are shown in Panel (a), (b), (c), and (d), respectively. In the absence of particle identification (PID) detector in the simulation, a $3\sigma$ separation power for PID is assumed in the prospected PID detector acceptance coverage listed in Table \ref{tab:pid_acceptance}. It is applied as a momentum hard cut-off as a function of $\eta$.

\begin{table}[b]
\caption{\label{tab:pid_acceptance}%
PID acceptance for hadrons at different $\eta$ regions.
}
\begin{ruledtabular}
\begin{tabular}{c|ccc}
$\eta$ &
$[-3,-1)$ &
$[-1,1)$ &
$(1,3]$ \\
\colrule
$p_{max}[\rm{GeV}/$$c]$  & 4 & 6 & 15\\
\end{tabular}
\end{ruledtabular}
\end{table}

\begin{figure}[htp]
  \centering
  \includegraphics[width=0.4\textwidth]{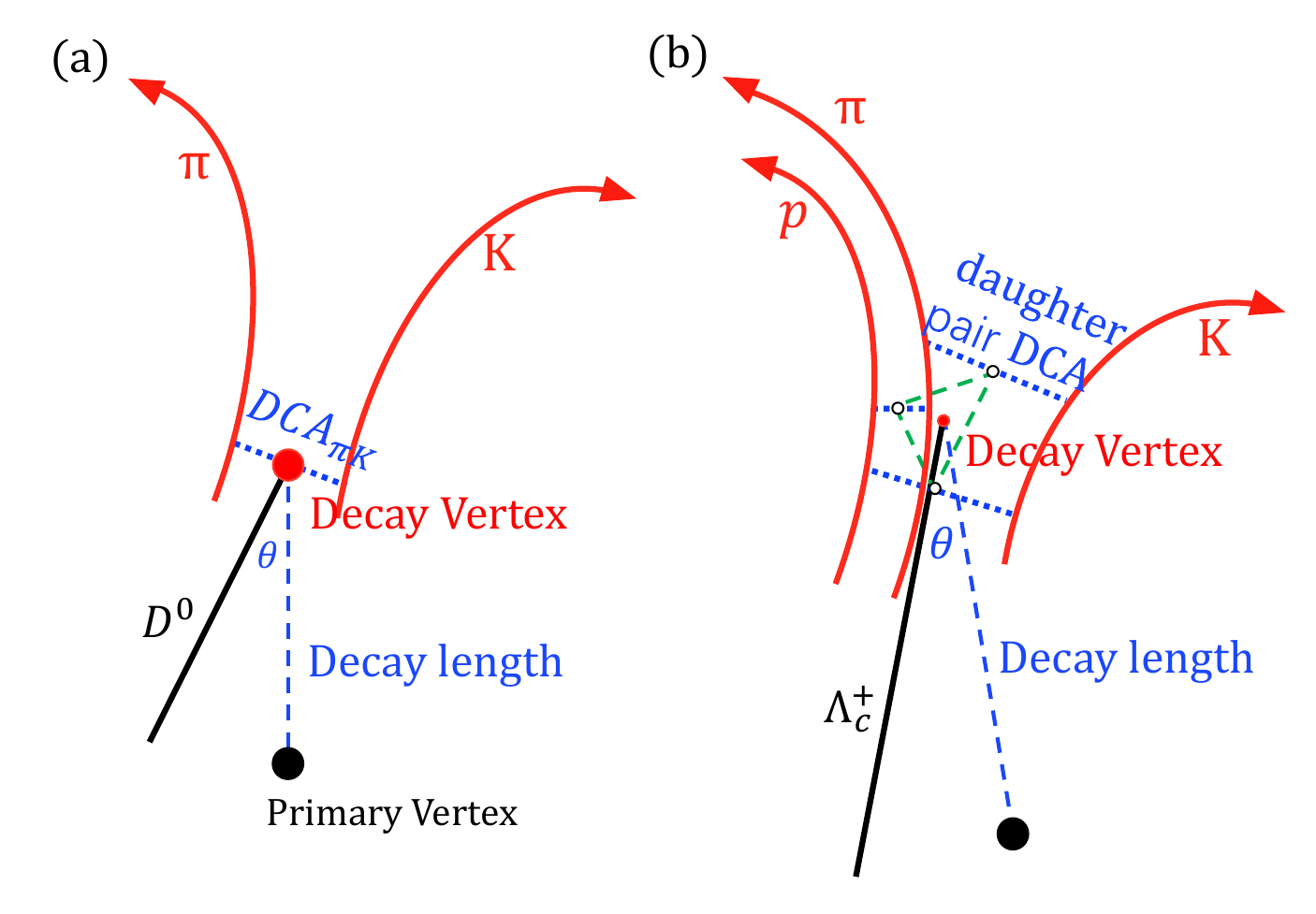}
  \caption{\label{fig:topological_cut_diagram} Topological cuts applied to reconstruct $D^{0}$ (a) and $\Lambda_{c}^+$ (b).}
\end{figure}

\begin{figure*}[htp]
  \centering
  \includegraphics[width=0.7\textwidth]{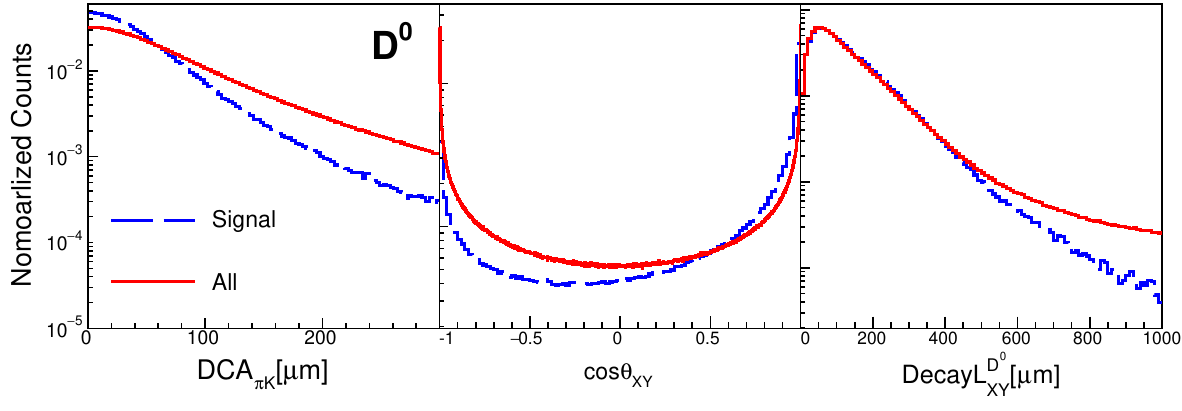}
  \caption{\label{fig:topological_cut_distribution}  Normalized $D^0$ Topological cut distributions, where the blue lines are for signal and the red lines are for background.}
\end{figure*}

All the performance parameters mentioned above have been used to conduct a fast simulation for the physical projection. In this simulation, the track information generated by \textsc{pythia} is adjusted, or ``smeared,'' according to a Gaussian function to mimic the limitations of the detection capability.
The number of the tracks in the acceptance ($-3<\eta<3$) and passing the tracking efficiency filters is used to determine the primary vertex resolution, which is also smeared in the \textsc{pythia} events. After smearing, different topological requirements are used to separate the signal and background. 

$D^0$ is reconstructed from channel $D^0\rightarrow\pi^+K^-$ with a branch ratio ($\mathcal{B}$) of 3.83$\%$ and $\Lambda_c^+$ is reconstructed from channel $\Lambda_c^+\rightarrow\pi^+K^-p$ ($\mathcal{B}$=2.96\% in \textsc{pythia 6.4} and $\mathcal{B}$=3.4\% in \textsc{pythia 8.3}). The $\mathcal{B}$ difference of $\Lambda_c^+$ between \textsc{pythia 6.4} and \textsc{pythia 8.3} is caused by a missing channel in \textsc{pythia 6.4} ($\Lambda_c^+\rightarrow\Lambda\pi^+, \Lambda\rightarrow pK^-$).  In the following part of this report, we use $D^0$ to represent both $D^0$ and $\bar{D^0}$ and use $\Lambda_c^+$ to represent both $\Lambda_c^+$ and $\Lambda_c^-$. Figure.~\ref{fig:topological_cut_diagram}, we show topological  cut diagrams for $D^0$ (a)  and for $\Lambda^{+}_c$ (b). For $D^0$, three topological requirements are applied, including $DCA_{\pi K}$, $DecayL_{XY}^{D^0}$ and $\cos \theta_{XY}$, which are defined as:
\noindent
\begin{itemize}
\item [$\bullet$] $DCA_{\pi K}$ is the closest distance between two daughter tracks.
\item [$\bullet$] $DecayL_{XY}^{D^0}$ is the distance between the decay vertex and the reconstructed primary vertex (in the $xy$ plane). Once $DCA_{\pi K}$ is defined, two points that symbolize $DCA_{\pi K}$ on each of the two tracks can be found. Then, the decay vertex of $D^0$ is defined as the middle of the two points.
\item [$\bullet$] $\cos \theta_{XY}$ is the cosine value of the angle between the $D^0$ decay length and the $D^0$ momentum (in the $xy$ plane).
\end{itemize}
The reason for defining both $\cos \theta_{XY}$ and $DecayL_{XY}^{D^0}$ in the $xy$ plane is that the spatial resolution in the $xy$ plane is much better than that in the $z$ dimension, especially in the endcap region. However, if the projections of two non-parallel lines in $xy$ plane cross, $DCA_{\pi K}$ will always be zero in the $xy$ plane. Therefore, $DCA_{\pi K}$ is defined in 3 dimensions instead of $xy$ plane. For $\Lambda_c^+$, the $DCA_{\pi K}$ is replaced by $DCA_{\rm daughters}^{\rm max}$, the maximum of all three $DCA_{\rm daughters}$, where $DCA_{\rm daughters}$ is the DCA between two of the three daughters. The decay vertex is the center of all three decay vertexes defined by each daughter pair. All three topological cut distributions are shown in Fig.\ref{fig:topological_cut_distribution} for the signal (blue dashed line) and background (red solid line). \par

\begin{figure*}[htp]
  \centering
  \includegraphics[width=0.7\textwidth]{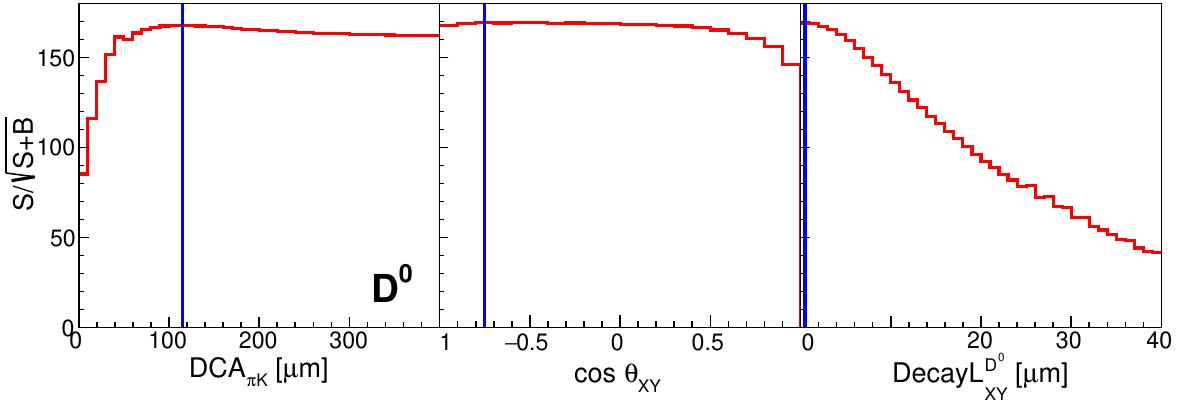}
  \caption{\label{fig:siginificance_cut_scan} Topological requirements are optimized to ensure the highest significance of $D^0$. The previous best results are applied to the next round of optimization.}
\end{figure*}

The $\pi K$ invariant mass distributions in the $D^0$ mass region are fitted with a Gaussian function for signal and a linear function for background to obtain the $D^0$ signal yield. 
The significance is defined as $S/\sqrt{S+B}$, where $S$ and $B$ stand for the counts of $D^0$ signal and background, respectively.
All the criteria are optimized by maximizing the significance iteratively.
Figure.~\ref{fig:siginificance_cut_scan} shows $D^0$ significance as a function of all three topological requirements. The optimal topological criteria are: (i) $DCA_{\pi K}<110\,\mu m$, (ii) $\cos \theta_{XY}>-0.75$. As seen in Fig.\ref{fig:d0_decay_kinematic} (the first panel), the $p_T$ of most $D^0$ is lower than 1GeV/$c$ and $D^0$ with lower $p_T$ tends to have a smaller $DecayL^{D^0}_{XY}$, so the $DecayL^{D^0}_{XY}$ distribution of signal is similar to that of background. Therefore, the significance is insensitive to $DecayL^{D^0}_{XY}$ as shown in Fig.\ref{fig:siginificance_cut_scan}. 

\begin{figure}[htp]
  \centering
  \includegraphics[width=0.4\textwidth]{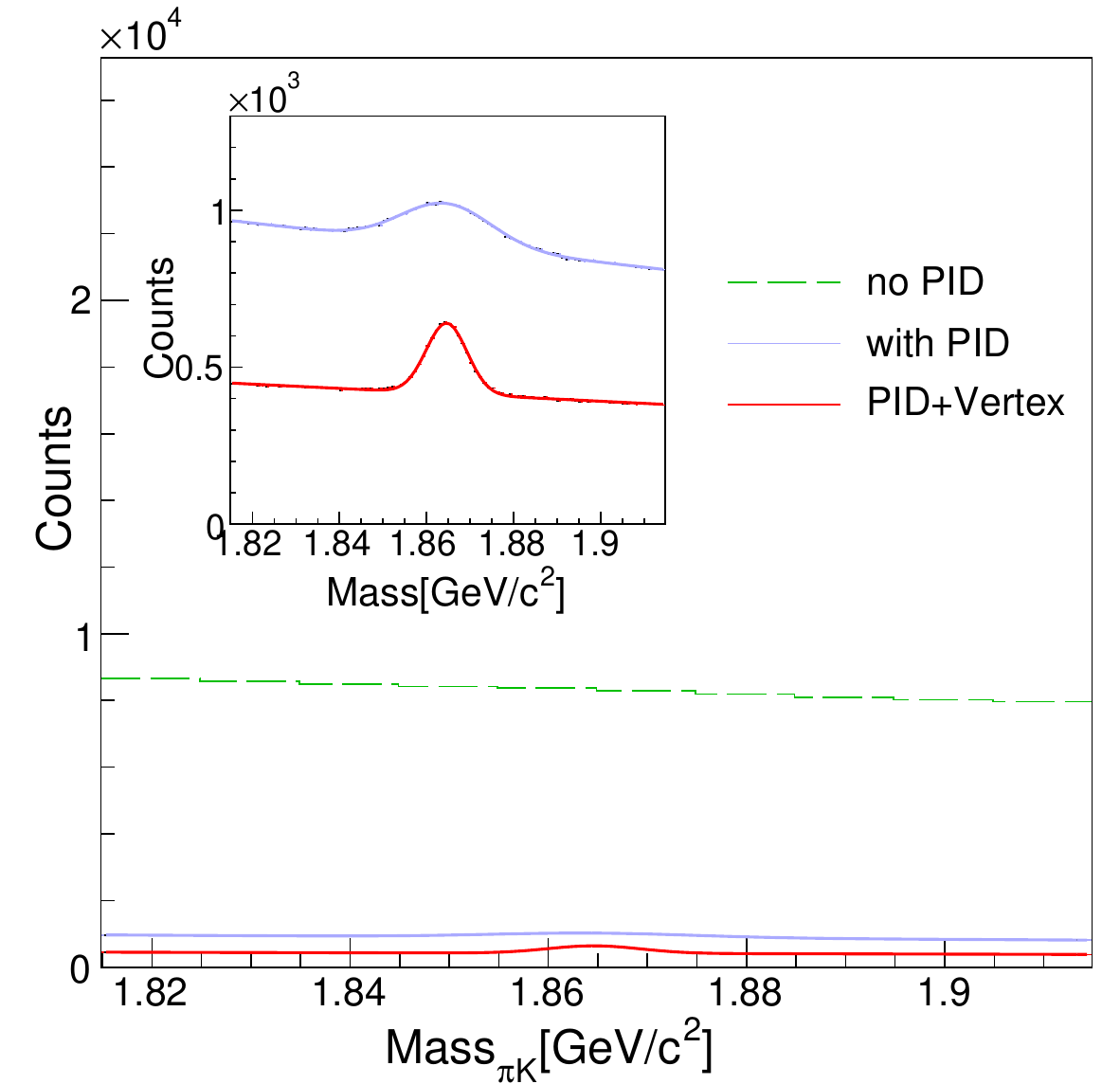}
  \caption{\label{fig:d0_reconstruction_comparison} The $\pi K$ invariant mass distributions with different detector performance configurations:($\textbf{no PID}$) without either PID system or vertex detectors, ($\textbf{with PID}$) with the PID system but without vertex detectors, ($\textbf{PID+Vertex}$) with the PID system and the vertex detectors. The significance is achieved at an integrated luminosity 0.04\rm{fb}$^{-1}$. The significances of $\textbf{with PID}$ and $\textbf{PID+Vertex}$ are 13.9 and 21.8, respectively. The $S/B$ of $\textbf{with PID}$ and $\textbf{PID+Vertex}$ are 0.06 and 0.215, respectively.}
\end{figure}

The $\pi K$ invariant mass distributions with different detector performance configurations are shown in Fig.\ref{fig:d0_reconstruction_comparison}. There are three configurations included: $\textbf{no PID}$, $\textbf{with PID}$ and $\textbf{PID+Vertex}$. 
In the first two configurations, the vertex detectors in the geometry for \textsc{geant4} simulation are removed to determine the significance of the vertex detectors in $D^0$ reconstruction.
For $\textbf{no PID}$, the PID acceptance in Table \ref{tab:pid_acceptance} isn't applied, and all possible combinations formed by particles with opposite charges are considered. 
As shown in Fig.\ref{fig:d0_reconstruction_comparison}, the signal peak of the green line ($\textbf{no PID}$) is almost invisible. 
Compared to the green line ($\textbf{no PID}$), the background of the blue line ($\textbf{with PID}$) is greatly suppressed when the PID acceptance is applied. 
Similarly, compared to the green line ($\textbf{no PID}$), the background of the red line ($\textbf{PID}\textbf{+}\textbf{Vertex}$) is also suppressed, and the mass peak is narrowed. This is because the momentum resolution becomes better when the vertex detectors are installed. After adding the vertex detectors, the significance improves from $13.9$ to $21.8$.

\begin{figure*}[htp]
  \centering
  \includegraphics[width=0.8\textwidth]{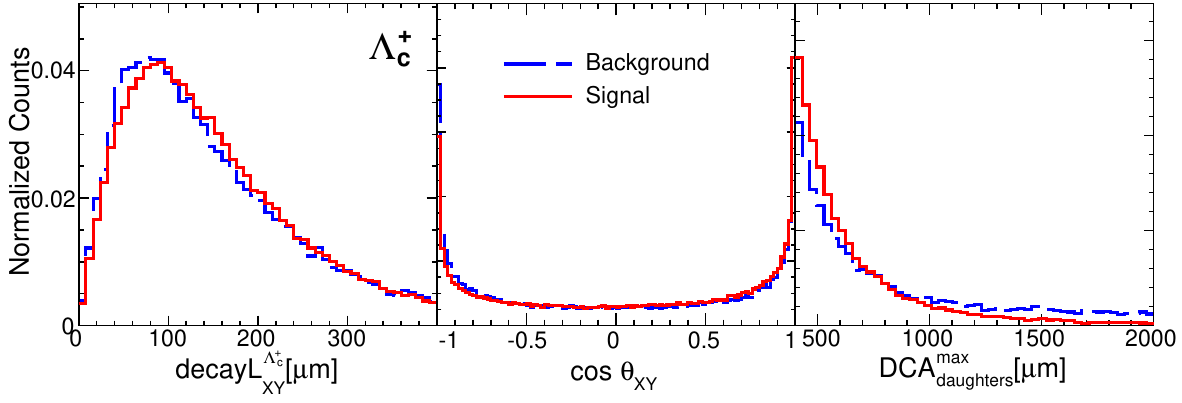}
  \caption{\label{fig:lc_cut_dis} $\Lambda_c^{+}$ topological cut distributions. The blue dashed and red solid lines represent the distributions for the background and signal, respectively. Because the differences of $decayL^{\Lambda^+_c}_{XY}$ and $\cos{\theta_{XY}}$ distribution between signal and background are  small, $DCA^{max}_{\rm Daughters}$ is the only topological requirement taken into account. }
\end{figure*}

\begin{figure}[htp]
  \centering
  \includegraphics[width=0.4\textwidth]{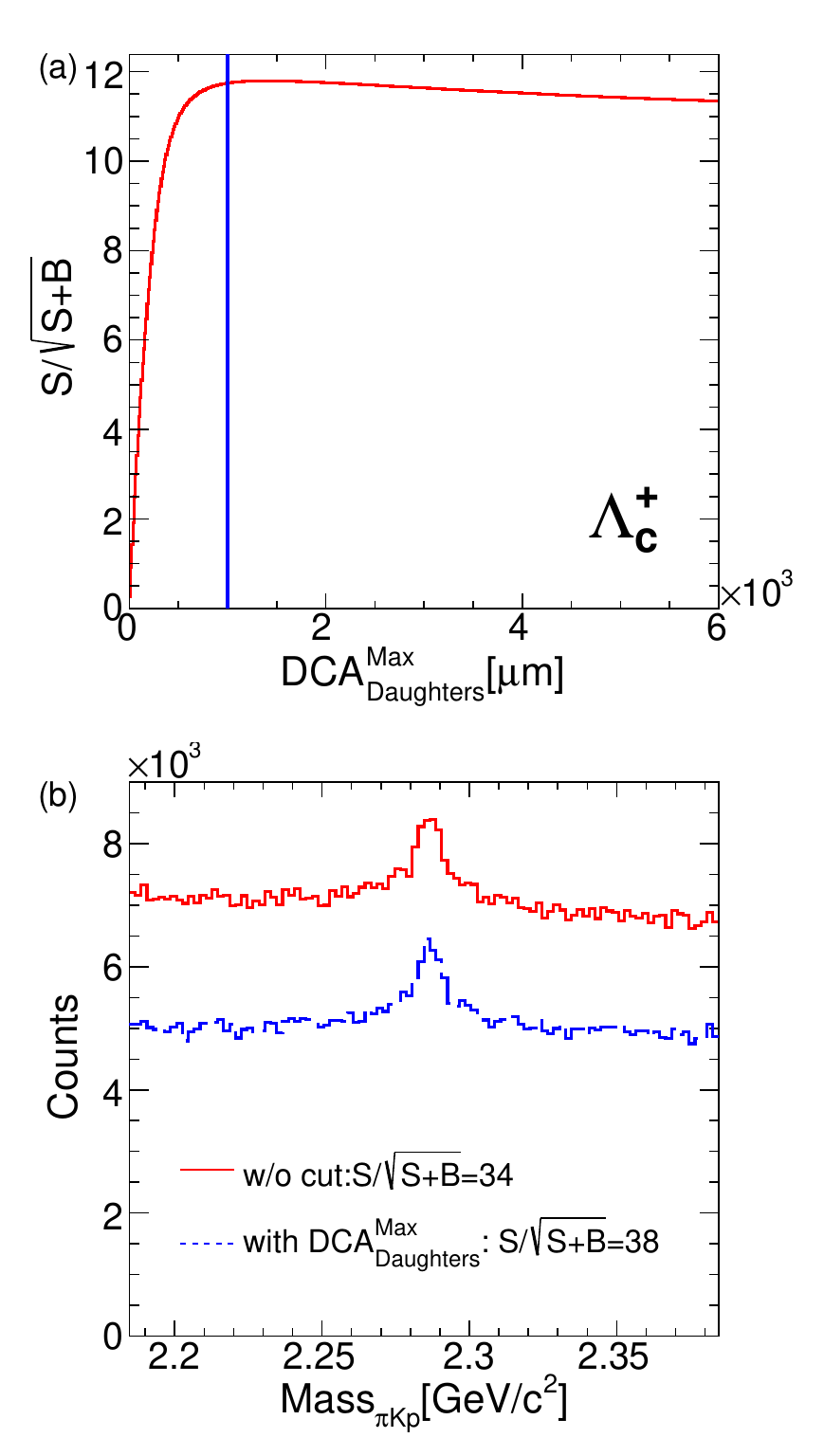}
  \caption{\label{fig:lc_mass_and_scan} The results of $\Lambda_c^{+}$ reconstruction.(a) the significance scanning for $DCA^{max}_{\rm Daughters}$. (b) the $\pi Kp$ invariant mass distributions. The read solid and blue dashed lines represent the distributions without and with requirement application, respectively.}
  \end{figure}

The reconstruction procedure is also carried out for $\Lambda_c^{+}$. The difference is that the shape of the $\Lambda_c^+$ peak is non-Gaussian, as shown in the bottom panel of Fig.\ref{fig:lc_mass_and_scan}.
Therefore, we use the line shape of $\Lambda_c^{+}$ after smearing the daughter tracks from pure $\Lambda_c^{+}$ and a linear function in the fit. The distributions of $\Lambda_c^+$ topological cuts are shown in Fig.\ref{fig:lc_cut_dis}, where we can find that $DCA_{\rm daughters}^{\rm max}$ is the only useful criterion. Therefore, in Fig.\ref{fig:lc_mass_and_scan}, only the criterion optimization for $DCA_{\rm daughters}^{\rm max}$ is performed and the optimal result is $DCA_{\rm daughters}^{\rm max}<1000\,\mu m$. The significance is improved from 34 to 38 with this requirement.

\section{Results}
\subsection{Baryon-to-meson ratios}

\begin{figure}[htp]
  \centering
  \includegraphics[width=0.4\textwidth]{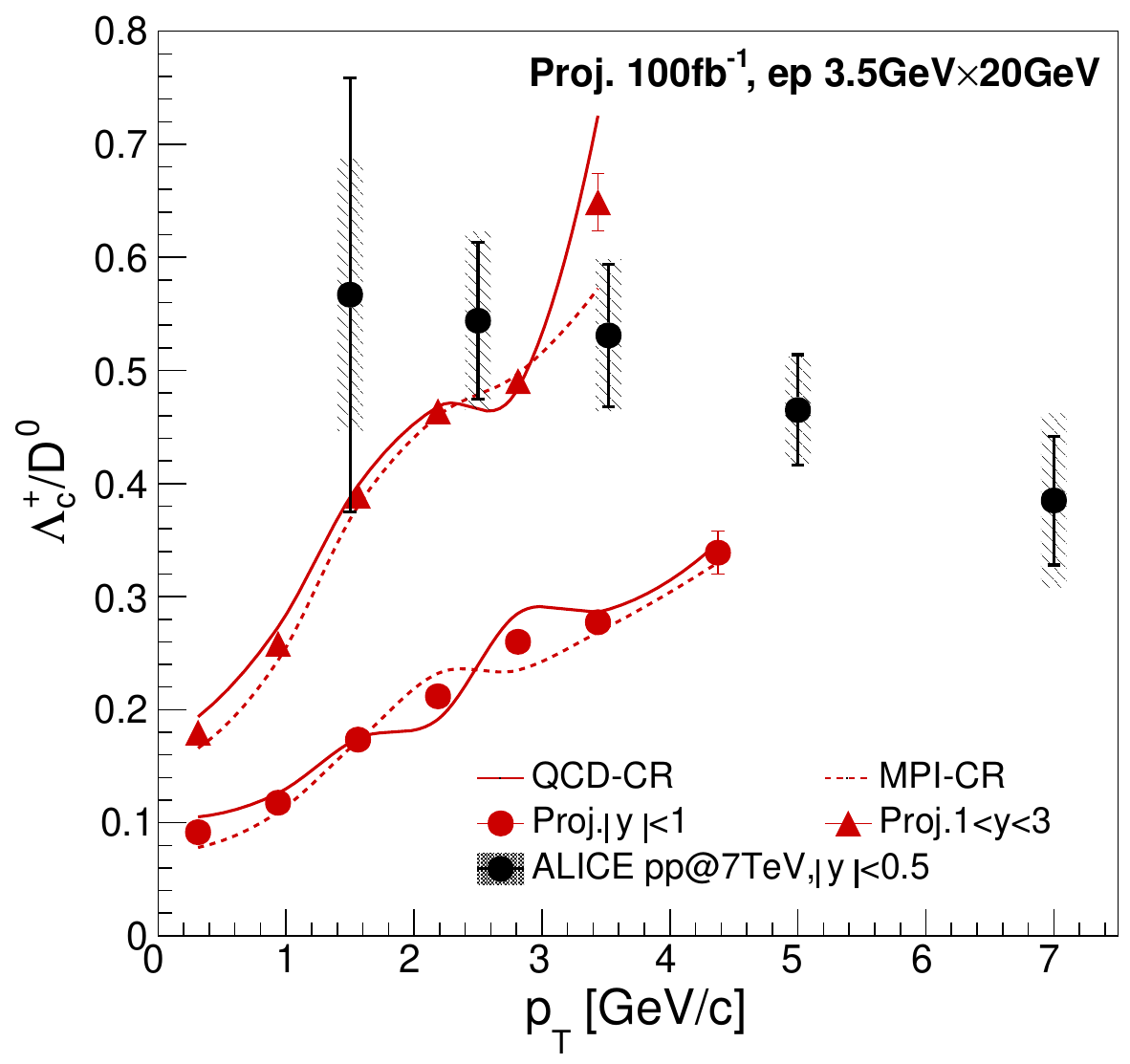}
  \caption{\label{fig:lc2d0_pt} The values of the red dashed and red solid lines are calculated from \textsc{pythia 8.3} with QCD-CR and MPI-CR. The red solid triangles and circles are the centers between the curves of the two models at both mid- and forward-rapidity. The red error bars represent the projection of $\Lambda_c^+/D^0$ ratio statistical uncertainties at an integrated luminosity $100\,\rm{fb}^{-1}$. The black points are the result of ALICE measurement where the error bars are statistical uncertainties and the grey boxes are the systematic uncertainties~\cite{lhc_frag_18}.}
\end{figure}

\begin{figure}[htp]
  \centering
  \includegraphics[width=0.4\textwidth]{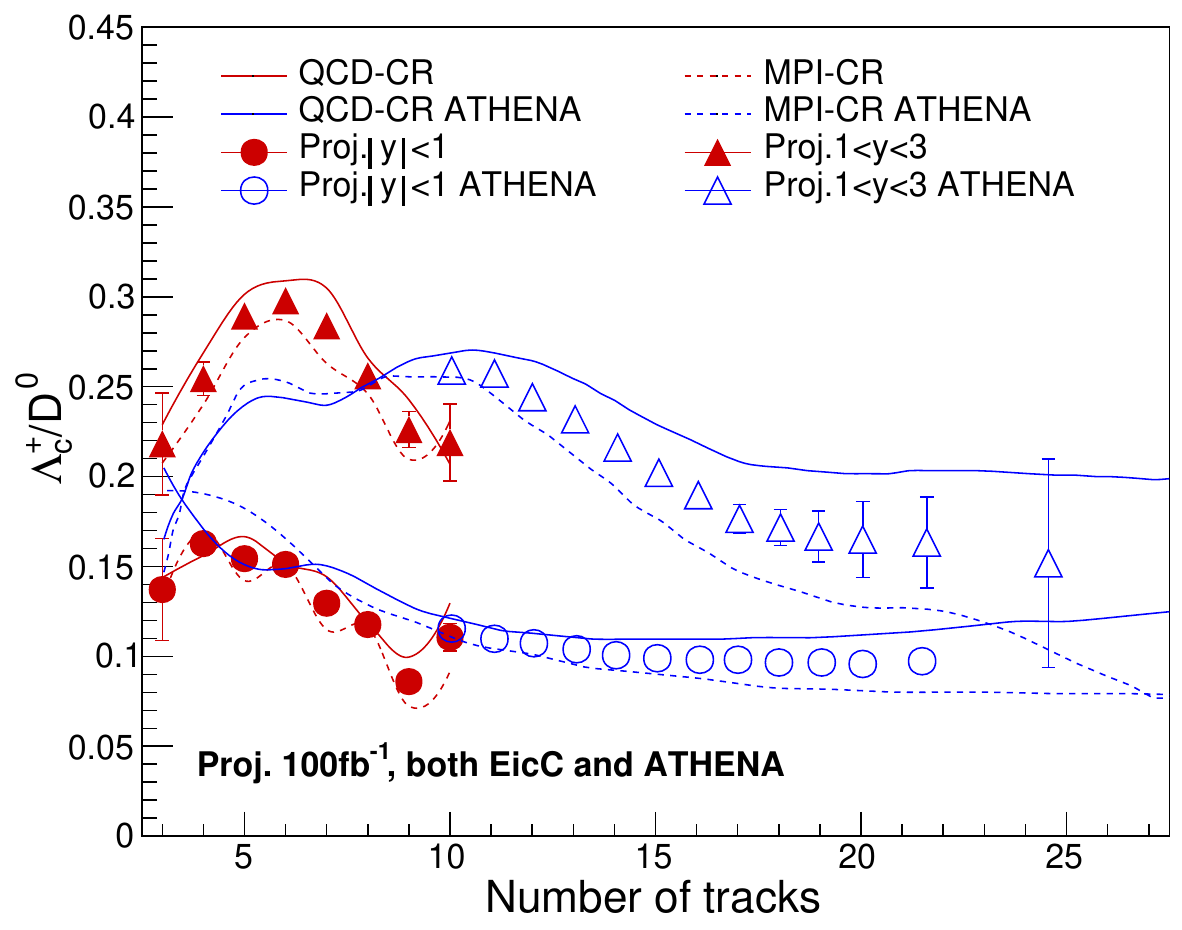}
  \caption{\label{fig:lc2d0_mult} The projection of the statistical uncertainty of $\Lambda_{c}^+/D^0$ ratio as a function of multiplicity. The description of both lines and points is similar to what has been discussed in Fig.\ref{fig:lc2d0_pt}. The red and blue points are for EicC results and ATHENA results~\cite{ATHENA_detector_proposal}.}
\end{figure}

\hspace{\parindent}The fragmentation ratios of the charm quark obtained from different experiments were thought to be universal. 
Both HERA $ep$ collision measurements~\cite{lc2d0_ratio_hera,lc2d0_ratio_hera_dis} and the previous $pp$ collision measurements at LHC~\cite{lhc_frag_2012} are consistent with $e^+e^-$ collision measurements~\cite{ee_c_frag}. However, an enhancement of $\Lambda_c^+/D^0$ ratio has been observed in the recent measurement at ALICE~\cite{lhc_frag_18,pp_frag_2021}.  The deviation of the $\Lambda_c^+/D^0$ ratio from $pp$ collision to $e^+e^-$ collision experiment indicates that even a small hadronic system can affect charm hadronization. 
Therefore, the $\Lambda_c^+/D^0$ ratio should be re-examined in the $ep$ collision system, specifically at EicC, which provides unique kinematic coverage and high luminosity.

Two projections of $\Lambda_c^+/D^0$ ratios are provided in this subsection. $ep$ collision data with beam energies 3.5 GeV $\times$ 20 GeV are generated by \textsc{pythia 8.3} with QCD color reconnection (QCD-CR) tuning~\cite{QCDCR} and multiple-parton-interaction color reconncetion (MPI-CR) tuning~\cite{beam_remnants}. In Fig.\ref{fig:lc2d0_pt}, the statistical uncertainty projection as a function of $p_T$ is shown. The curves are from \textsc{pythia 8.3}, in which the solid and dashed curves are from the QCD-CR and MPI-CR settings. The values of the red points are the average of the two models at the corresponding rapidity regions.  The estimation of statistical uncertainties is the root mean square of the sum of squares of errors propagated from all sources. For $\Lambda_c^+/D^0$ ratio, the expression is $$\sigma(\frac{\Lambda_c^+}{D^0})=\sqrt{(\frac{\sigma(\Lambda_c^+)}{N_{D^0}})^2+(\frac{N_{\Lambda_c^+}}{N^2_{D^0}}\sigma(D^0))^2}$$, where the statistical uncertainties of $D^0$ and $\Lambda_c^+$($\sigma(D^0)$ and $\sigma(\Lambda_c^+)$) are calculated by the signal yields divided by the corresponding significance. The statistical uncertainty is scaled to an integral luminosity $100\,\rm{fb}^{-1}$.
The results of $\Lambda_c^+/D^0$ ratios are presented in two rapidity regions, the mid-rapidity region ($|y|<1$) and forward-rapidity region ($1<y<3$), shown as solid circles and triangles respectively.  The measurement in the $pp$ collision from ALICE~\cite{lhc_frag_18} is also presented, and shown as black solid circles in Fig.\ref{fig:lc2d0_pt}. The black points represent the center values, the black line is the statistical uncertainty, and the grey box is the systematic uncertainty. Compared to ALICE, the statistical precision is significantly improved in EicC due to high luminosity and much less hadronic combinatorial background. EicC results show that the uncertainties are smaller than the difference between the two models at both mid- and forward-rapidity. This suggests that EicC has sufficient power to distinguish between different hadronization models. In addition, the $\Lambda_c^+/D^0$ ratios have different behaviors in mid- and forward-rapidity regions. A wide rapidity coverage, benefiting from large acceptance of the detector design, can be achieved at EicC. As mentioned above, there is an enhancement of $\Lambda_c^+$ production caused by beam remnant formation in the very forward rapidity region. With wider rapidity coverage of EicC, the interaction between beam remnant and charm quark can also be studied in greater detail. 

Fig.\ref{fig:lc2d0_mult} shows the statistical uncertainty projections of $\Lambda_c^+/D^0$ as a function of charged particle multiplicity by red marks. The high multiplicity results from EIC-BNL simulation~\cite{ATHENA_detector_proposal} are shown as the open blue marks. Only the tracks in $|\eta|<3$, $p_T>0.2\,\rm{GeV}/$$c$ and not from $D^0$ or $\Lambda_c^+$ decay are counted in the number of charge tracks. The dashed and solid lines are from the calculation of MPI-CR and QCD-CR \textsc{pythia 8.3} models, respectively. The red and blue curves are the results for EicC and EIC-BNL, respectively. EicC can provide a measurement of $\Lambda_c^+/D^0$ ratio at low multiplicity as a complement to EIC-BNL measurements. The measurement of the charm baryon-to-meson ratio can provide insight into the interplay between the hard and soft processes that produce particles. In $pp$ or $AA$ collisions, a higher charged particle multiplicity corresponds to higher collision energy and lower impact parameters. The events with different collision energies and impact parameters will have different underlying event structures. Thus, a wider coverage of multiplicity is necessary to study the charm hadronization at $\gamma-Nucleon$ and $\gamma-Nuclei$  collision systems. 

\subsection{$D^0$ double ratio}

\begin{figure}[htp]
  \centering
  \includegraphics[width=0.4\textwidth]{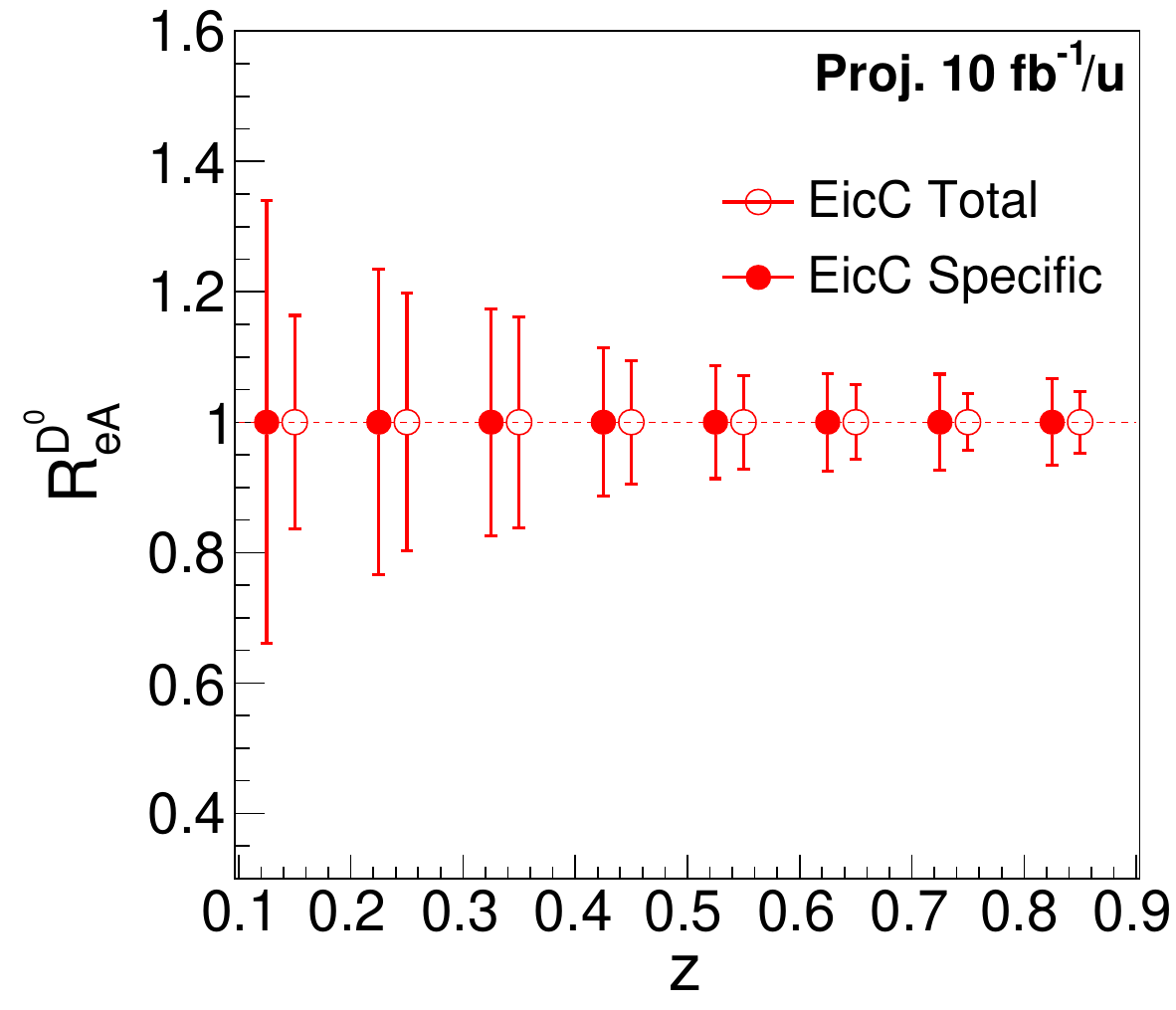}
  \caption{\label{fig:D0_double_ratio}The $D^0$ double ratio: The statistical uncertainty projections are corresponding to an integral luminosity $\mathcal{L}_{int}=10\,\rm{fb}^{-1}$. The open and solid circles represent the results of EicC total kinematic coverage (EicC Total: $2\,\rm{GeV}^2<Q^2<200\,\rm{GeV^2}, \nu<96\,\rm{GeV}$) and a smaller specific kinematic coverage (EicC Specific: $2\,\rm{GeV}^2<Q^2<8\,\rm{GeV}^2$, $48\,\rm{GeV}< \nu<77\,\rm{GeV}$) .}
\end{figure}

\hspace{\parindent}Hadron production is affected by initial and final state interactions. The primary initial effect is EMC effect, which is not of interest in this particular study. 
We focus on the final state interaction and the effect of the surrounding nuclear medium on hadronization. To minimize the initial state effect, the observable double ratio is defined as
$$R^{h}_{eA}(Q^2,x_B,z)=\frac{N^{h}_{A}(Q^2,x_B,z)/N^{DIS}_{A}(Q^2,x_B)}{N^{h}_{p}(Q^2,x_B,z)/N^{DIS}_{p}(Q^2,x_B)}$$
where $N^{DIS}$ and $N^{h}$ are the number of DIS events and the number of the SIDIS events in which $h$ hadrons are produced, respectively. 

DIS requirements, including (i) $Q^2>2\,$GeV$^2$, (ii) $\nu<96\,\rm{GeV}$, and (iii) $W^2>4\,$GeV$^2$ are applied. The statistical uncertainty projection of $D^0$ double ratio as a function of $z$ is shown in Fig.\ref{fig:D0_double_ratio}. Here, the electron-nucleus collision is the $eAu$ collision with electron momentum $3.5\,\rm{GeV}/c$ and $Au$  momentum $12.93\,\rm{GeV}/$$c$$/u$.  The center values of all the points are set to 1 because there should be no medium modification in \textsc{pythia}. The statistical uncertainty is scaled to the integral luminosity $10\,\rm{fb}^{-1}/u$ of electron-nucleus and electron-proton collision data. The open red circles are EicC results for the whole kinematic region. The solid red circles are EicC results for additional DIS criteria: $Q^2<8$\,GeV$^2$ and $48$\,GeV $<\nu<77$\,GeV. The EicC specific kinematic coverage in Fig.\ref{fig:D0_double_ratio} is chosen to be a region where $\nu$ is rather low and the $D^0$ are produced abundantly.

There are precise measurements of the light hadron double ratio ($\pi^{+/-}, K^{+/-}, p$ and $\bar{p}$) versus multiple variables at HERMES. However, as mentioned earlier, two theories with different time scales can adequately describe the HERMES data~\cite{parton_energy_loss,hadron_absorption}.
To provide an additional powerful constraint to these theories, $D^0$ double ratio is necessary because that $D^0$ double ratio is very different from that of light hadrons. For light hadrons, the values of the double ratio at all $z$ bins are below unity for nuclear attenuation or parton energy loss. However, for $D^0$, there is a peak in  $c\rightarrow D^0$ fragmentation function at low $z$($z<0.1$). So $D^0$ double ratio is higher than unity at low $z$ bins and lower than unity at high $z$ bins for the same reason as light hadron suppression. In addition, a lower c.m. energy can produce a larger cold nuclear effect~\cite{LI2021136261}. 
Therefore, EicC, which has a lower c.m. energy than EIC-BNL but can abundantly produce $D^0$, can be an ideal facility to study how the nuclear medium affects the hadronization process. Additionally, it can complement EIC-BNL by covering a wider kinematic region.

\section{Summary}


In summary, open charm reconstruction is studied with fast simulation, where the detector performance parameterizations are derived from the \textsc{geant4} simulation with the developing detector geometry designed for EicC. The reconstruction of $D^0(\rightarrow K^-\pi^+)$ and $\Lambda_c^+(\rightarrow p\pi^+K^-)$ in the $\sqrt{s}=16.7\,$GeV$\,ep$ collision demonstrates the importance of vertex detectors in reducing background and improving momentum resolutions. This leads to higher signal significance and lower statistical uncertainties, providing sufficient precision for future charm hadronization studies in EicC. 

We have projected the statistical uncertainty of $\Lambda_c^+/D^0$ ratios as functions of $p_T$ and charged-particle multiplicity. Different hadronization models (QCD-CR and MPI-CR) can be distinguished with the projected statistical uncertainties at an integrated luminosity of $\mathcal{L}_{int}=100\,\rm{fb}^{-1}$. A broader rapidity coverage than ALICE can provide abundant information about hadronization in $\gamma-Nucleon$ and $\gamma-Nuclei$ collision systems.

The statistical uncertainties of the double ratio of $D^0$, which behave differently from light hadron double ratios, are projected as a function of z at an integrated luminosity of $\mathcal{L}_{int}=10\,\rm{fb}^{-1}/u$. Precise measurements at EicC can provide an excellent opportunity to understand charm hadronization mechanisms and how charm interacts with the nuclear medium.

\acknowledgments{
This work is supported in part by the Strategic Priority Research Program of the Chinese Academy of Sciences under grant number XDB34000000, the Guangdong Major Project of Basic and Applied Basic Research No. 2020B0301030008, the Guangdong Provincial Key Laboratory of Nuclear Science with No. 2019B121203010, the National Natural Science Foundation of China with Grant No. 11890712, 12061141008 and the National Key R\&D Program of China with Grant No. 2018YFE0104700 and 2018YFE0205200. The authors acknowledge the computing resources available at the Southern Nuclear Science Computing Center. }

\bibliography{apssamp}

\end{document}